\title{Geographically Weighted Regression Analysis for Spatial Economics Data:
a Bayesian Recourse}
\date{}
\author{Zhihua Ma~~~~Yishu Xue~~~~Guanyu Hu}

\documentclass[12pt]{article}
\usepackage{setspace}
\usepackage[margin=1in, lines=25]{geometry} 
\usepackage{graphicx}
\usepackage{subcaption}
\usepackage{natbib,caption,color,bm}
\usepackage{xr-hyper}
\RequirePackage[colorlinks, citecolor=blue, linkcolor=blue]{hyperref}
\newcommand{\blue}[1]{{\textcolor{black}{#1}}}

\newcommand{\cov}{\textrm{cov}}
\usepackage{mathrsfs,physics}
\usepackage{float}
\usepackage{amsmath}
\usepackage{booktabs}
\usepackage{multirow}
\usepackage{threeparttable}
\usepackage{listings}
\renewcommand{\hat}{\widehat}
\renewcommand{\bar}{\overline}
\newcommand{\proglang}[1]{\textsf{#1}}
\newcommand{\pkg}[1]{{\normalfont\fontseries{b}\selectfont #1}}
\newcommand{\IGamma}{\textrm{IGamma}}

\newcommand{\ACC}{\text{ACC}}

\begin{document}
\maketitle
\begin{abstract}
The geographically weighted regression (GWR) is a well-known statistical
approach to explore spatial non-stationarity of the regression relationship in
spatial data analysis. In this paper, we discuss a Bayesian recourse of GWR.
Bayesian variable selection based on spike-and-slab prior, bandwidth selection
based on range prior, and model assessment using a modified deviance information
criterion and a modified logarithm of pseudo-marginal likelihood are fully
discussed in this paper. Usage of the graph distance in modeling areal data is
also introduced. Extensive simulation studies are carried out to examine the
empirical performance of the proposed methods with both small and large number
of location scenarios, and comparison with the classical frequentist GWR is
made. The performance of variable selection and estimation of the proposed
methodology under different circumstances are satisfactory. We further apply the
proposed methodology in analysis of a province-level macroeconomic data of 30
selected provinces in China. The estimation and variable selection results
reveal insights about China's economy that are convincing and agree with
previous studies and facts.
\bigskip

\end{abstract}
\newpage
\section*{Introduction}
For geographically sparse data with inherent spatial variability, estimating
coefficients of a regression model for a particular location based on only
observations from this location is not feasible due to the small number of
observations. The geographically weighted regression
\citep[GWR;][]{brunsdon1996geographically}
is an important tool to explore spatial non-stationarity of the regression
relationship in spatial data analysis. It has been applied to a variety of
fields, including geology, environmental science, epidemiology, and
econometrics. \cite{fotheringham2003geographically} has summarized the basic
theory, statistical inference, and bandwidth selection for GWR, and proposed
natural extensions of GWR under the generalized linear model framework.
The basic idea of GWR is to make use of information from nearby locations. The
weighting idea is a natural strategy to use in light of Tobler's first law that
``near things are more related than distant things'' \citep{tobler1970computer}.
In estimating the parameter for one specific location, subjects in the data are
weighed according to their distance from this location, with greater weight for
closer subjects. \cite{paez2002general1,paez2002general2} proposed estimation
and inference for GWR under a maximum-likelihood-based framework.
\cite{mei2006testing} proposed a mixed geographically weighted regression model
which included both spatially-varying and non-spatially-varying coefficients,
and gave a testing procedure of important explanatory variables.
\cite{wang2008local} proposed a local linear-based GWR for the spatially varying
coefficient models which can significantly improve GWR. More recently,
\cite{da2016multiple} discussed the multiple testing issue, and proposed a
solution which outperforms other solutions such as the Bonferroni procedure
under the GWR framework. The aforementioned works discussed the GWR in the
frequentist fashion. From the Bayesian perspective, \cite{LeSage2004} proposed a
Bayesian GWR, which gives a prior distribution on the parameter vector depending
on historical knowledge. The proposed model, however, used cross-validation for
bandwidth selection, which relies on a user-specified grid of bandwidth, and is
computationally intensive like other cross-validation based methods.

In this paper, we propose Bayesian techniques for the GWR using the
likelihood-based approach in \cite{paez2002general1,paez2002general2}. In
addition to Bayesian estimation and inference, a spike and slab
\citep{ishwaran2005spike} prior is applied for variable selection for
Bayesian GWR. Furthermore, bandwidth selection and weighting scheme selection
are discussed based on prior selection and Bayesian model selection criteria.
An introduction to the implementation of GWR based on \pkg{nimble}
\citep{de2017programming}, a relatively new and powerful \proglang{R} package
for Bayesian inference, is presented as an open source repository on GitHub.
Our simulation studies showed the promising empirical
performance of the proposed methods in both non-spatially varying and spatially
varying cases. In addition, our proposed Bayesian approach reveals interesting
features of the province-level macroeconomic data in China.

The rest of this paper is organized as follows. In Section ``Geographically
Weighted Regression",
the GWR and its weighting schemes are discussed. Section ``Bayesian Recourse for
GWR" gives a detailed discussion of Bayesian inference, variable selection,
bandwidth selection, and model assessment for the GWR. Extensive simulation
studies are conducted in Section ``Simulation Studies" to investigate the
empirical performance of the proposed methods. In Section ``Real Data Analysis",
we implement our model using  province-level macroeconomic data in China from
year 2012 to year 2016. Finally, Section ``Discussion" contains a brief summary
of this paper.

\section*{Geographically Weighted Regression }\label{Geographically weighted
regression}
\subsection*{Geographically Weighted Regression}

From \cite{brunsdon1998geographically} , the GWR
model can be written as:
\begin{eqnarray}\label{eq:gwr}
y(s)=\beta_1(s)x_1(s)+...+\beta_p(s) x_p(s)+\epsilon(s)
\end{eqnarray}
where $y(s)$ is the response variable at location $s$, $\beta_i(s)$,
$i=1,2,...,p$ are the coefficients of independent variables at location $s$, and
$\epsilon(s)$ is the random effect at location $s$, assumed to follow
$N(0,\sigma^2)$. In addition, we also assume
$\cov(\epsilon(\ell),\epsilon(m))=0$ for any $\ell\neq m$. Given a weighting
function, the weights of each observation can be calculated with the distance
between that observation and~$s$. Estimation of coefficients at location~$s$ can
be formulated in a way similar to the weighted least squares:
\begin{eqnarray}\label{eq:freqGWR}
\hat{\beta}(s)=\left(X^\top W(s)X\right)^{-1}X^\top W(s)Y,
\end{eqnarray} 
where $X$ is the $n\times p$ matrix of covariates, $Y$ is the $n\times 1$ vector
of responses, and $W(s) = \textrm{diag}(w_1(s),...,w_n(s))$ is a diagonal matrix
of the weights.

\subsection*{Spatial Weighting Functions and
Distances}\label{sec:weightingfunction}

We first introduce several spatial weighting functions that can be used in GWR.
Notice that the weighting scheme of ordinary least squares can be defined in the
following form:
\begin{eqnarray}
w_{jk}=1,~~  \forall j,k
\end{eqnarray}
where $j$ represents the location of the observations, and $k$ represents the
location for which parameters are estimated. In a global model where
observations from all locations are used to estimate one vector of coefficients,
each observation is assigned a weight of unity.

A first step to consider locality is to include observations that are
only within a certain distance~$d$ from the target location, i.e.,
\begin{equation}\label{eq:stepWeight}
	w_{jk} = \begin{cases}
		1 & d_{jk}\leq d\\
        0 & \text{otherwise}
	\end{cases},
\end{equation}
where~$d_{jk}$ is the distance between locations~$j$ and~$k$. This weighting
scheme is one of the simplest to calculate. It is, however, a discontinuous
function of distance, which can sometimes lead to undesired jumps in the
estimated parameter surface. In order to get a continuous weighting function,
the exponential function and the Gaussian function can also be used. The
exponential weighting scheme can be written as:
\begin{eqnarray}\label{eq:expweight}
w_{jk}=\exp\left(-d_{jk}/b\right),
\end{eqnarray}
where $b$ is the bandwidth that can be chosen appropriately to control the decay
with respect to distance.
The Gaussian weighting scheme can be written as:
\begin{eqnarray}\label{eq:gaussianweight}
w_{jk}=\exp\left(-(d_{jk}/b)^2\right).
\end{eqnarray}
Both \eqref{eq:expweight} and \eqref{eq:gaussianweight} are decreasing functions
of $d_{jk}$, which, intuitively, indicates that an observation very far away
from the location of interest contributes little in the estimation of parameters
at this location. In order to provide a continuous, near-Gaussian weighting
function up to distance $b$ from the estimation point, and then zero weights for
any data point beyond $b$,
\cite{brunsdon1996geographically,brunsdon1998geographically,fotheringham1998geographically}
proposed the bi-square function:
\begin{equation}
	w_{jk} = \begin{cases}
		(1-(d_{jk}/b)^2)^2 & d_{jk}<b\\
	0 & \text{otherwise}
	\end{cases}.
\end{equation}
For the bi-square kernel, by tuning the threshold $b$, one can control the
number of neighbors that are used to estimate the parameters for the location of
interest. The weighting schemes mentioned above are the most popular schemes
used in the GWR.

We then briefly discuss different choices of distance function. The Euclidean
distance, defined~as
\begin{equation*}
d_{jk} = \sqrt{(\text{latitude}_j - \text{latitude}_k)^2 + 
(\text{longitude}_j - \text{longitude}_k)^2},
\end{equation*}
is one of the most popular choices when the precise (latitude, longitude)
location of each observation is available. However, in some public health and
epidemiology studies, or some socioeconomics studies, data are collected and
summarized on a higher level than single observations, such as
wards~\citep{brunsdon1996geographically} or
counties~\citep{xue2018geographically}, which produces areal data instead of
point-reference data. All observations within the same area are assigned the
same (latitude, longitude). For example, \cite{hu2018modified} attributed each
county's observations to its centroid. Note that the Euclidean distance is
easily affected by the areas of the administrative divisions, which additionally
complicates the process of parameter tuning as there is no golden benchmark
measure of distance.

An alternative distance measure when we have areal data is the graph distance
\citep{muller1987algorithm,bhattacharyya2014community}. The administrative
devisions are regarded as vertices of a graph $G$, denoted as $v_1,\ldots, v_n$.
The graph $G$ also includes a set of edges, $E(G) = \{e_1,\ldots, e_m\}$, where
each edge connects a pair of vertices. The graph distance is defined as:
\begin{equation}
	d_{v_i v_j} = \begin{cases}
        \vert V(e) \vert & \text{ if $e$ is the shortest path connecting
        $v_i$ and $v_j$}
\\
		\infty & \text{if $v_i$ and $v_j$ are not connected}
	\end{cases},
\end{equation}
where $\vert V(e) \vert$ denotes the number of edges in $e$.

While it remains a subjective problem in choosing appropriate bandwidths and
thresholds for the geographical distance based methods, i.e. one has to decide
``how close is close enough", a natural definition of closeness would derive
from the graph distance. Counties sharing a common boundary, i.e. having graph
distance 1, are close, while having graph distance greater than 1 indicates
``not close", and observations in these far neighboring counties need to be
weighed down. A graph distance based weighting function would be
\begin{equation}\label{eq:GDistWeight}
    w_{jk} = \begin{cases}
    1 & d^G_{jk} \leq 1 \\
    f(d^G_{jk}\mid b) & \text{otherwise}
    \end{cases},
\end{equation}
where $d_{jk}^G$ denotes the graph distance, and $f$ is a certain weighting
function with bandwidth $b$. \blue{In this work, we choose $f()$ to be
the negative exponential function, i.e.,
\begin{equation}\label{eq:GDistExp}
    w_{jk} = \begin{cases}
        1 & d_{jk}^G \leq 1 \\
        \exp(-d_{jk}^G / b) & \mbox{otherwise}
    \end{cases}.
\end{equation}}

\section*{Bayesian Recourse for GWR}\label{sec:bayesian_gwr}

In this section, we propose the posterior estimation, variable selection, and
bandwidth selection for the Bayesian GWR model. The proposed methods are
implemented with the powerful \proglang{R} package \pkg{nimble}. The code and
documentation can be found at GitHub.

\subsection*{Bayesian Estimation for GWR}
According to \cite{Boscardin94bayesiancomputation} and
\cite{paez2002general1,paez2002general2}, we can get the estimation of GWR using
Bayesian computation. The likelihood function of this model can be written as:
\begin{eqnarray}
Y \mid \beta(s),X,W(s),\sigma^2(s) \sim \mbox{MVN}(X\beta(s),\sigma^2(s) W^{-1}(s)),
\end{eqnarray}
where MVN indicates the multivariate normal distribution. In order to have a
conjugate posterior distribution, we can set the priors of $\beta(s)$ and
$\sigma^2(s)$ as:
\begin{eqnarray}
\beta(s) \mid \Sigma_\beta \sim N_p(0,\Sigma_\beta),
\end{eqnarray}
where $\Sigma_\beta$ is a diagonal matrix, and
\begin{eqnarray}
\sigma^2(s) \sim \IGamma(\alpha_1,\alpha_2),~~j=1,\ldots,p,
\end{eqnarray}
where $\alpha_1,\alpha_2$ are the hyper-parameters for
distributions of  $\sigma^2(s)$. One set of
non-informative choices of hyper-parameters
is to set $\Sigma_\beta=100I_p$ and $\alpha_1=\alpha_2=0.01$
\citep{gelman2013bayesian}.
The posterior distribution can be written as:
\begin{eqnarray}\label{eq:posterior}
p\left(\beta(s),\sigma^2(s)\mid
Y,X,W(s)\right)\propto p\left(Y\mid \beta(s),X,W(s),
\sigma^2(s)\right)\times
p\left(\beta(s)\mid \Sigma_\beta\right)\times p\left(\sigma^2(s)\right).
\end{eqnarray}
According to \eqref{eq:posterior}, we can use Markov chain Monte Carlo
\citep[MCMC,][]{gelman2013bayesian}
to estimate~$\beta(s)$ and $\sigma^2(s)$.

\subsection*{Bayesian Variable Selection}

We first consider the regression problem for one location. Following the
procedure of \cite{george1993variable}, the spike and slab prior for
$\beta_j(s)$ can be formulated as:
\begin{eqnarray}
\beta_j(s)\mid \gamma_j\sim (1-\gamma_j)N(0,\tau_j^2)+\gamma_jN(0,c_j^2\tau_j^2),
\end{eqnarray}
where
\begin{eqnarray}
P(\gamma_j=0)=1-P(\gamma_j=1)=p_j.
\end{eqnarray}
When $\gamma_j=0$, $\beta_j(s) \sim N(0,\tau_j^2)$, and when $\gamma_j=1$,
$\beta_j(s)\sim N(0,c_j^2\tau_j^2)$. Our interpretation of this prior is: we set
$\tau_j$ small enough so that if $\tau_j=0$, $\beta_j(s)$ would be so small that
we can ``safely'' estimate it as 0; inversely, we set $c_j$ large so that if
$\gamma_j=1$, we include the $\beta_j(s)$ into our final model. For the prior of
$\gamma_j$, we set $\gamma_j \sim \text{Bernoulli}(0.5)$, which is a
non-informative choice.

\subsection*{Bayesian Bandwidth Selection}

In GWR, it is important to choose a proper bandwidth for the weighting
functions. In the Bayesian approach, a prior can be given to the bandwidth $b$
so that the optimal bandwidth can be simultaneously obtained together with the
estimation of other parameters. The prior also depends on which measure of
distance is used. A more detailed discussion of distance measures is given in
Section ``Spatial Weighting Function and Distances''. Using similar ideas as in
\cite{boehm2015spatial}, a prior for bandwidth can be set as:
\begin{equation*}
	b \sim \text{Uniform}(0,D),
\end{equation*}
where $D$ is the upper limit for the support of the distribution of $b$. Without
any prior knowledge,~$D$ can be chosen large enough so that we start from a
noninformative prior for the bandwidth, i.e., we start from an approximate
global model where observations are always weighed equally. There are also some
other choices of prior distributions for the bandwidth, such as the gamma
distribution or discrete uniform distribution. If prior information is available
about the bandwidth, parameters for the prior distributions can be set to
incorporate such information. Our proposed model can be summarized as follows:
\begin{eqnarray}
& Y\mid \beta(s),X,W(s),\sigma^2(s) \sim \mbox{MVN}(X \beta(s),\sigma^2(s)W^{-1}(s))
\label{eq:Ymodel}\\ \ &\beta_j(s)\mid \gamma_j, \tau_j\sim
(1-\gamma_j)N(0,\tau_j^2)+\gamma_jN(0,c_j^2\tau_j^2)\label{eq:betamodel} \\
	&\tau_j^2 \sim \text{IGamma}(\alpha_1,\alpha_2) \\
	&\gamma_j \sim \text{Bernoulli}(0.5) \\
	&w_i(s)=f(d_i\mid b) \\
	&b\sim \text{Uniform}(0, D)\label{eq:bmodel}
\end{eqnarray}
where ``IGamma" denotes the inverse-Gamma distribution, and $f$ is the weighting
function introduced in \eqref{eq:GDistWeight}, $i=1,...,n$ and $j=1,...,p$. As
we incorporate the prior of $b$ into our model, conjugate posterior distribution
for $b$ cannot be obtained. Therefore, we use the Metropolis--Hastings Algorithm
\citep[MH;][]{gelman2013bayesian} to estimate the parameters.

\subsection*{Bayesian Model Assesment}
In Section ``Spatial Weighting Function and Distances'', we introduced several
spatial weighting functions that can be used in the GWR. In order to select the
weighting scheme that fits the data best, we apply the most commonly used tools,
the Deviance Information Criterion \citep[DIC;][]{spiegelhalter2002bayesian} and
the Logarithm of the Pseudo-Marginal Likelihood
\citep[LPML;][]{ibrahim2013bayesian}, for model selection. The DIC is defined
as:
\begin{gather}\label{formula:DIC}
	\text{DIC} = \text{Dev}(\bar{\theta}) + 2 p_D,
\end{gather}
where $\theta$ and $\bar{\theta}$ represent the parameter of interest and the
corresponding posterior mean. The term~$\text{Dev}(\cdot)$ denotes the deviance
function, while $p_D$ is the effective number of parameters in the model, given
by $p_D = \overline{\text{Dev}}(\theta) - \text{Dev}(\bar{\theta})$. For the GWR
model in our paper, the following deviance function can be specified
\citep{ma2018bayesian}:
\begin{align*}
& \text{Dev}(\beta(s), W(s), \sigma^2(s)) = -2
\text{log}f(Y\mid \beta(s), X, W(s), \sigma^2(s)) \\
=~&n \text{log}(2\pi) + \text{log}(\sigma^2(s)) - \text{log}(\left| W(s)\right |)
+ (Y - X\beta(s))^\top \sigma^{-2}(s) W(s) (Y - X\beta(s)),
\end{align*}
where $n $ is the total number of the observations. Therefore, the DIC for the
GWR model can be given as:
\begin{align*}
\text{DIC} &= \text{Dev}(\bar{\beta}(s), \bar{W}(s), \bar{\sigma}^2(s)) +
2p_D\\
&= 2 \overline{\text{Dev}}(\beta(s), W(s), \sigma^2(s)) -
\text{Dev}(\bar{\beta}(s), \bar{W}(s), \bar{\sigma}^2(s)),
\end{align*}
where $\bar{\beta}(s)$, $\bar{W}(s)$ and $\bar{\sigma}^2(s)$ are posterior
estimates obtained from MCMC results. A smaller value of DIC indicates a better
model. It can be regarded as the Bayesian equivalent of AIC, where the term
$p_D$ is the penalty term for model complexity, similar to the $p$, i.e.,
dimension of parameter space, in AIC. Similar to AIC, DIC also takes both
fitness and model complexity into account simultaneously.

The LPML is constructed based on the Conditional Predictive Ordinate (CPO)
values, which are estimates of the probability for observing $Y_i$ given that
all other responses have been observed. Let $D_{(-i)} = \{Y_j: j = 1, \cdots,
i-1, i+1, \cdots, N\}$ denote the observed data with the $i$th subject response
deleted. The CPO for the $i$th subject is defined as:
\begin{gather}
\text{CPO}_i = \int f(Y\mid \beta(s), X, W(s), \sigma^2(s)) \pi(W(s), \beta(s),
\sigma^2(s)\mid D_{(-i)}) \dd (W(s), \beta(s), \sigma^2(s)),
\end{gather}
where $\pi(W(s), \beta(s), \sigma^2(s)\mid D_{(-i)}) = \frac{\prod_{j\neq
i}f(Y_j\mid \beta(s), X,W(s), \sigma^2(s)) \pi(W(s), \beta(s), \sigma^2(s)\mid
D_{(-i)})}{c(D_{(-i)})}$, and $c(D_{(-i)})$ denotes the normalizing constant.
Within the Bayesian framework, a Monte Carlo estimate of the CPO can be obtained
as:
\begin{gather*}
\widehat{\text{CPO}}_i^{-1} = \frac{1}{T} \sum_{t=1}^{T} \frac{1}{f(Y_i\mid X_i,
\beta_t(s), W_t(s), \sigma_t^2(s))},
\end{gather*}
where $T$ is the total number of Monte Carlo iterations. Then an estimate of the
LPML is given by:
\begin{gather}
	\widehat{\text{LPML}} = \sum_{i=1}^{N} \text{log}(\widehat{\text{CPO}}_i).
\end{gather}
A model with a larger LPML value indicates that it is more preferred. 

\section*{Simulation Studies}\label{sec:simu}

\begin{figure}
    \centering
    \begin{subfigure}{.7\textwidth}
        \includegraphics[width=\textwidth]{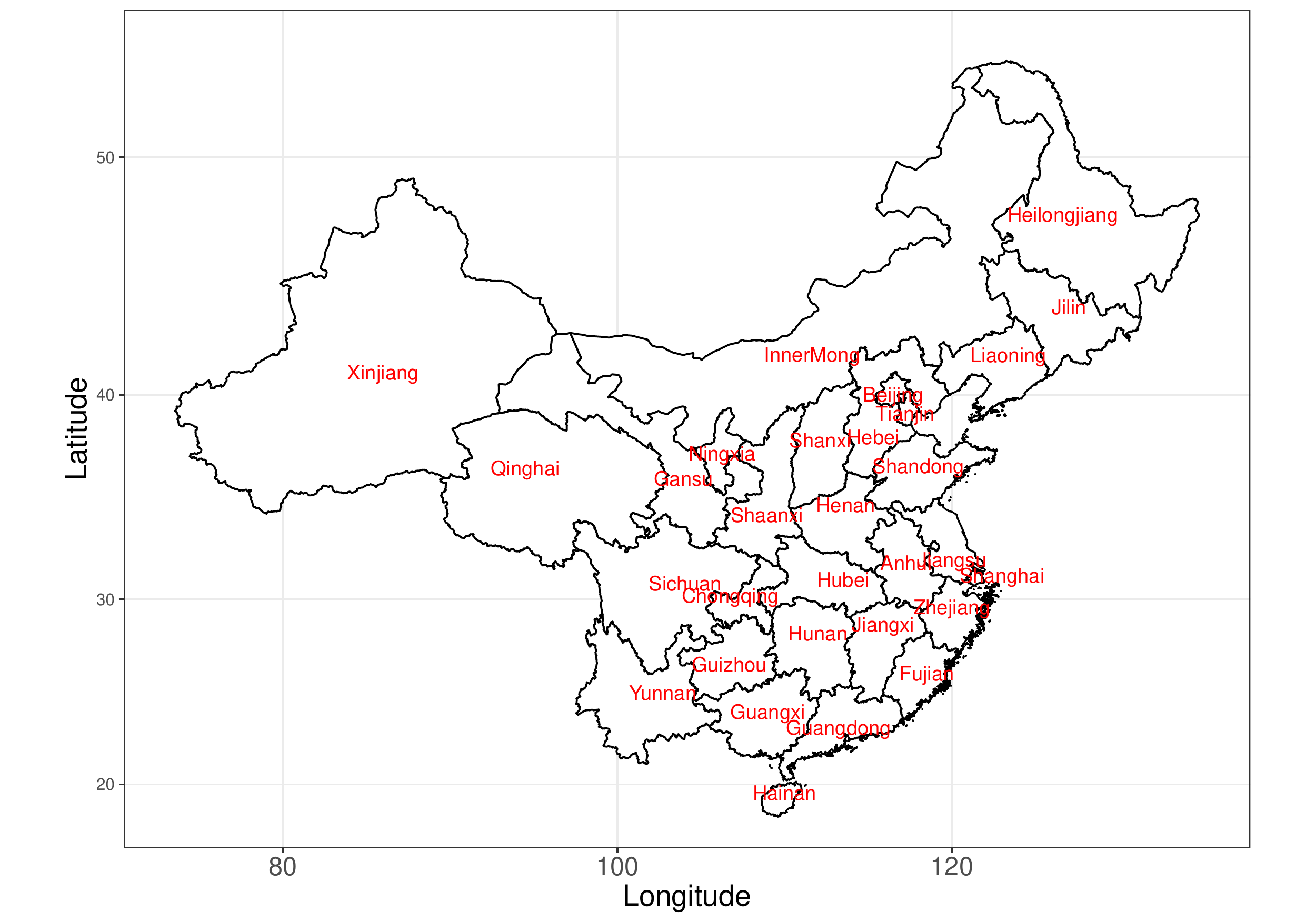}
        \caption{}
        \label{fig:mapwithname}
    \end{subfigure}
    \begin{subfigure}{.7\textwidth}
        \includegraphics[width=\textwidth]{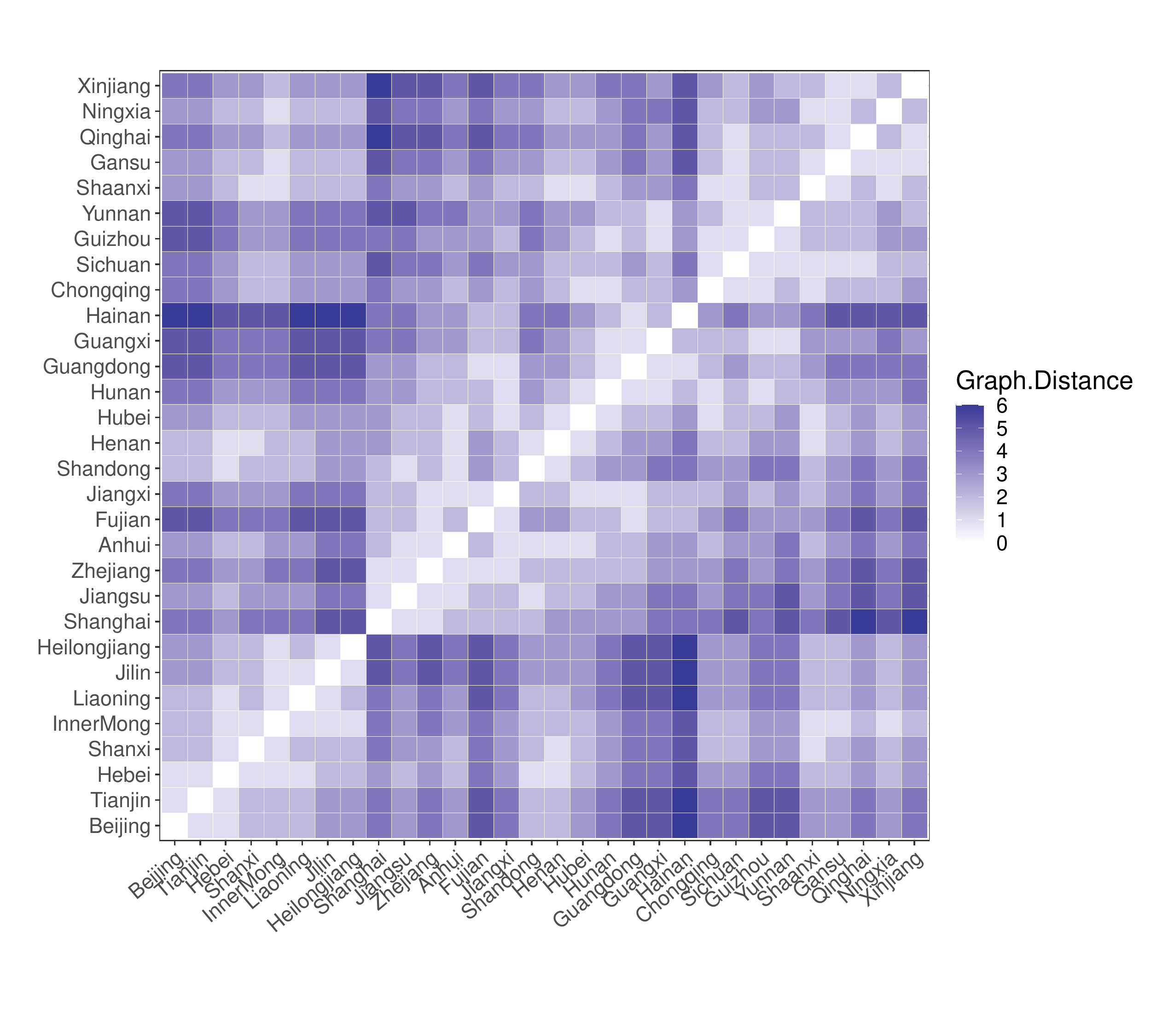}
        \caption{}
        \label{fig:gdist}
    \end{subfigure}
    \caption{(a) A map of the 30 selected provinces of China, with their names
    annotated. (b) Visualization of graph distance matrix for 30
    selected provinces in China. Darker color indicate larger graph distance.}
        \label{fig:fig1}
\end{figure}

In this section, we present the performance of the proposed estimation and
variable selection techniques under scenarios where the covariate effects do not
vary spatially, and where the covariate effects do vary spatially. We use the
spatial structure of 30 selected provinces in China in our simulations. A map of
these provinces with their names is presented in Figure~\ref{fig:fig1}(a).
Specifically, Hainan province is an island, and is therefore not connected with
any others, which makes its graph distance with any other province infinity. It
is, however, very close to Guangdong province, and they bear a lot of
resemblance in both culture and economic development. Therefore, we modified the
adjacency matrix so that Hainan and Guangdong are adjacent. The graph distance
matrix is calculated based on the modified adjacency matrix. A visualization of
the graph distance matrix is presented in Figure~\ref{fig:fig1}(b).

Denote the average parameter estimates as $\bar{\beta}_{\ell,m}$, calculated as 
\begin{equation}\label{eq:avgbeta}
\bar{\hat{\beta}}_{\ell,m} = \frac{1}{100}\sum_{r=1}^{100}\hat{\beta}_{\ell,
    m,r},
\end{equation}
where $\hat{\beta}_{\ell, m, r}$ denotes the parameter estimate for the $m$th
coefficient of province $\ell$ in the $r$th replicate. The parameter estimates
are evaluated based on their bias, standard deviation, mean squared error, and
coverage rate of the 95\% highest posterior density (HPD) intervals in the
following four ways:
\begin{align}
    \text{mean absolute bias (MAB)} & = \frac{1}{30}\sum_{\ell=1}^{30}
    \frac{1}{100} \sum_{r=1}^{100} \left|\hat{\beta}_{\ell, m,r} - 
    \beta_{\ell,m} \right|,  \label{eq:mab}\\
    \text{mean standard deviation (MSD)} & = \frac{1}{30}\sum_{\ell=1}^{30}
    \sqrt{\frac{1}{99}\sum_{r=1}^{100} \left(\hat{\beta}_{\ell, m, r} -
    \bar{\hat{\beta}}_{\ell,m}\right)^2},  \label{eq:msd}\\
    \text{mean of mean squared error (MMSE)} & = \frac{1}{30}\sum_{\ell=1}^{30}
    \frac{1}{100}\sum_{r=1}^{100} \left(\hat{\beta}_{\ell,m,r} - \beta_{\ell,m}
    \right)^2,  \label{eq:mmse} \\
    \text{mean coverage rate (MCR)} & = \frac{1}{30}\sum_{\ell=1}^{30}
    \frac{1}{100}\sum_{r=1}^{100} \text{I}\left(
    \beta_{\ell, m} \in 95\% \text{ HPD interval}
    \right), \label{eq:mcr}
\end{align}
where $\beta_{\ell,m}$ is the true underlying parameter, and $\text{I}(\cdot)$
denotes the indicator function. These measures are first calculated for each
individual province over replicates, and then averaged over provinces. The
variable selection approach is evaluated using both the accuracy rate for a
single variable ACC and for the entire model (Model $\ACC$), defined as:
\begin{align*}
\text{ACC}_m = \begin{cases}
\frac{1}{100}\sum_{r=1}^{100} \text{I (covariate $m$ is selected in the final
model)} & \beta_{\ell,m} \neq 0,~\ell = 1,\ldots,30 \\
\frac{1}{100}\sum_{r=1}^{100} \text{I (covariate $m$ is not selected in the
final model)} & \beta_{\ell,m} = 0, ~\ell = 1,\ldots, 30
\end{cases},
\end{align*}
and
\begin{align*}
    \text{Model ACC} = \frac{1}{100}\sum_{r=1}^{100}\text{I (the exact true
    underlying model is selected)}.
\end{align*}
To compare with frequentist approach, the same datasets are also fitted using
the classical frequentist GWR approach, where the bandwidth selection is made
based on minimizing the summation of SSE across 30 provinces over a grid of
candidate bandwidths. The details of frequentist bandwdith selection, as well as
the parameter estimates obtained using the optimal bandwdith, are presented in
Section~1 of the supplemental material.
\blue{To demonstrate
that the graph distance produces credible parameter estimates, and that at the
same time it circumvents the additional effort of threshold selection
in weighting kernels such as \eqref{eq:stepWeight} and \eqref{eq:GDistExp},
simulation study is done for the same designs to be presented, with the great
circle distance used. The results are reported in
Section~2 of the supplemental material.}
In addition, considering that a total of 150 observations
with~30 locations still make a small sample, an additional simulation study with
more than 300 locations using the spatial structure of census tracts in
Hartford, Litchfield, and Middletown counties in Connecticut has been conducted,
and included in Section~3 of the supplemental material.

For both the following simulation studies and the supplemental simulation
studies, the effective number of parameters for the frequentist GWR is also
calculated as in \cite{brunsdon2000geographically}. Note that the frequentist
GWR is based on one bandwidth only, and only the full model that includes all
covariates is fitted, therefore the frequentist $p_D$ should be used as a
reference, instead of a criteria for direct comparison.

\subsection*{Simulation Without Spatially Varying Coefficients}

Under the scenario where there are no spatially varying coefficients, we
generate data using the same set of parameters for all provinces. The
independent continuous covariates are generated i.i.d. from the standard normal
distribution $\mathcal{N}(0,1)$, denoted as $X_1$, $X_2$,\ldots, $X_5$, and we
use the matrix $X$ to denote the covariate matrix with 5 columns, with the $i$th
column being $X_i$. The response vector $Y$ is generated as $X \beta +
\bm{\epsilon}$, where $\epsilon\sim \mbox{MVN}(\bm{0},\bm{I})$. Different
choices of $\beta$ have been used corresponding to different underlying true
models. The parameter $D$ for bandwidth is set to be~100. Given that the maximum
graph distance in the spatial structure of the 30~selected provinces is~6, a
bandwidth of 100 induces a weighting schemes that, even if the distance between
one certain province and another province whose parameter estimates we want to
obtain, this province gets assigned a relative weight of $\exp(-6/100)$, which
is approximately~0.941 and thus approximates a global model where every
observation is equally weighed. This ensures that the prior for bandwidth~$b$ is
sufficiently noninformative. For each province, five observations are generated,
resulting in~150 observations per replicate. A total of 100 replicates are
performed. For each replicate, a chain of length 10,000 is run without thinning,
where the first 2,000 samples are discarded as burn-in. Three parameter settings
similar to in \cite{Shao:1997wn} were used, with $\beta^\top = (2,0,0,4,8)$,
$(2, 2, 0, 4, 8)$, and $(2, 2, 3, 4, 8)$, respectively. The mean of bandwidths
selected in the 100 replicates was also calculated. The results are reported in
Table~\ref{tab:simnull}.

It is rather clear that when there is no spatial variation, the bandwidth is
selected to be large, which induces a weighting scheme that assigns close to
uniform weight to both nearby provinces and distant provinces. Under all three
settings, the variable selection accuracies are all 100\% for all five
covariates, and the three model selection accuracies are 100\% as well. The
average effective number of parameters under the frequentist GWR are 11.08,
10.27 and 12.65 under the three settings, while under the Bayesian GWR, the
values are 178.00, 177.89 and 177.57, respectively.

\begin{table}[tbp]
\centering
\caption{Average parameter estimates, performance of parameter estimates, and
variable selection results when there is no spatial variation in the underlying
true parameters. `` True" denotes whether a covariate is in the true model or
not, and ACC stands for variable selection accuracy rate.}\label{tab:simnull}
    \begin{tabular}{ccccccccccccc}
    \toprule
& & $\bar{\hat{\beta}}$&MAB & MSD & MMSE & MCR & True &
ACC(\%) & Model ACC(\%) & $b$ \\
    \midrule 
    Setting 1 & $\beta_1$ & 1.998 & 0.055 & 0.074 & 0.005 & 0.942 & 1 & 100 & 100 &
    70.112\\
    & $\beta_2$ & 0.007 & 0.037 & 0.046 & 0.002 & 0.990 & 0 & 100 \\
    & $\beta_3$ & 0.004 & 0.040 & 0.049 & 0.002 & 0.999 & 0 & 100  \\
    & $\beta_4$ & 4.019 & 0.062 & 0.077 & 0.006 & 0.974 & 1 & 100 \\
    & $\beta_5$ & 7.990 & 0.066 & 0.085 & 0.007 & 0.934 & 1 & 100 \\ [0.5ex]
    Setting 2 & $\beta_1$ & 1.998 & 0.055 & 0.073 & 0.005 & 0.942 & 1 & 100 & 100 &
    70.213\\
    & $\beta_2$ & 2.012 & 0.064 & 0.080 & 0.006 & 0.975 & 1 & 100\\
    & $\beta_3$ & 0.003 & 0.040 & 0.049 & 0.002 & 1.000 & 0 & 100\\
    & $\beta_4$ & 4.019 & 0.062 & 0.077 & 0.006 & 0.975 & 1 & 100\\
    & $\beta_5$ & 7.991 & 0.066 & 0.085 & 0.007 & 0.934 & 1 & 100\\ [0.5ex]
    Setting 3 & $\beta_1$ & 1.997 & 0.055 & 0.073 & 0.005 & 0.939 & 1 & 100 &
    100 & 70.058\\
    & $\beta_2$ & 2.012 & 0.065 & 0.080 & 0.007 & 0.973 & 1 & 100\\
    & $\beta_3$ & 3.005 & 0.068 & 0.084 & 0.007 & 0.954 & 1 & 100\\
    & $\beta_4$ & 4.018 & 0.062 & 0.076 & 0.006 & 0.977 & 1 & 100\\
    & $\beta_5$ & 7.990 & 0.067 & 0.085 & 0.007 & 0.936 & 1 & 100\\
    \bottomrule
    \end{tabular}
\end{table}

\subsection*{Simulation with Spatially Varying Coefficients}

\begin{figure}[tbp]
\centering
\includegraphics[width = \textwidth]{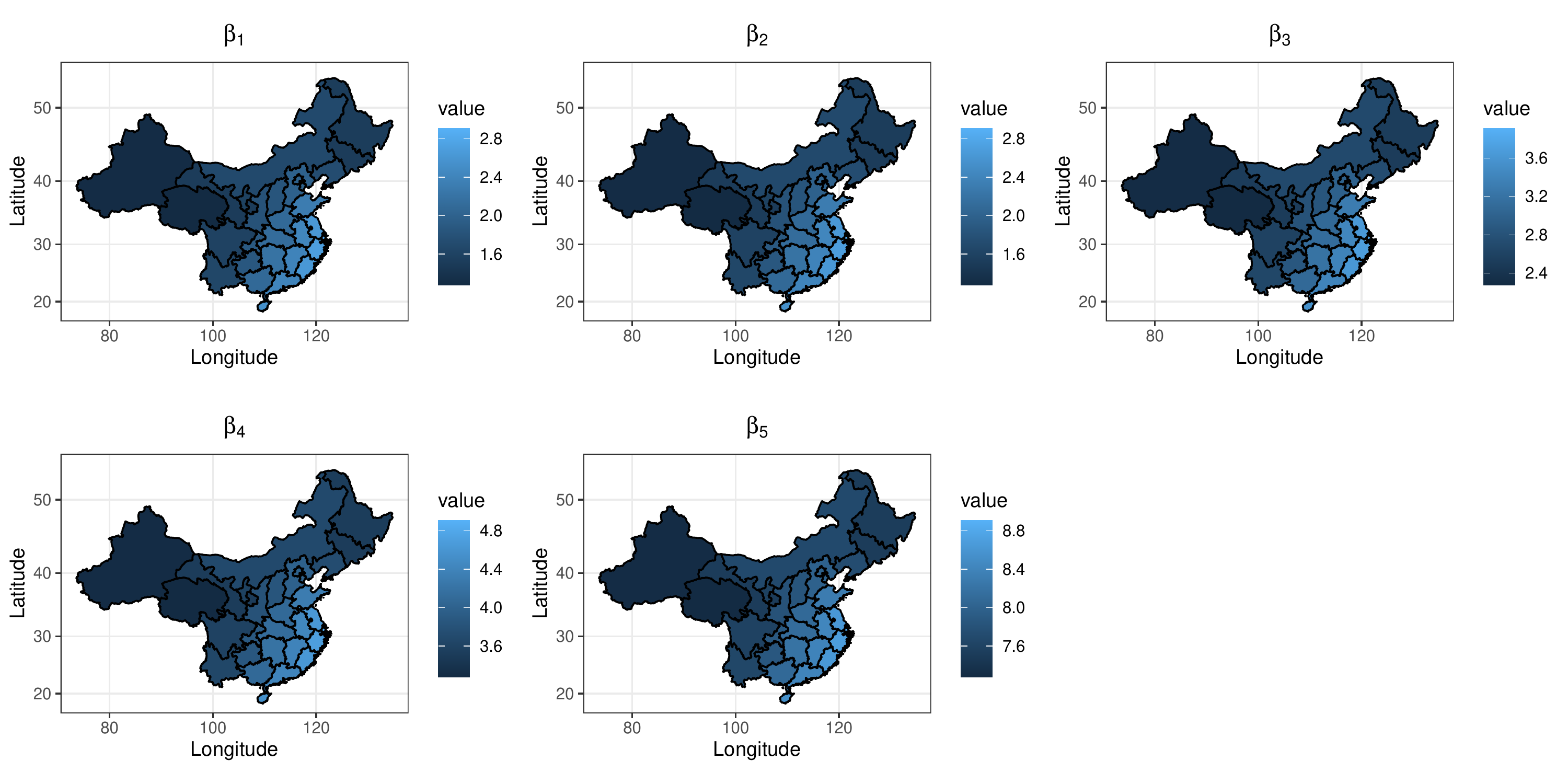}
\caption{Visualization of parameter surfaces when the parameters vary
according to \eqref{eq:alter1}.}\label{fig:parsurfacealter1}
\end{figure}

For estimation and variable selection in the presence of spatially varying
coefficients, we use similar simulation schemes as
in~\cite{xue2018geographically}. The graph distance matrix visualized in
Figure~\ref{fig:fig1} is transformed using multidimensional scaling
\citep[MDS;][]{cox2000multidimensional} and mapped into a Cartesian space.
Denoting the transformed coordinates corresponding to province $\ell$ as
$(x^c_\ell, y^c_\ell)$, the $\beta$ vector for province $\ell$ is set to
\begin{equation}\label{eq:alter1}
    \beta_{p,\ell} = \begin{cases}
        0, &\text{$x_p$ not in the true model}\\
        \beta_p + 0.2(x_\ell^c + y_\ell^c),&\text{$x_p$ in
the true model}
    \end{cases}.
\end{equation}
The variation pattern is visualized in Figure~\ref{fig:parsurfacealter1}. The
estimation and variable selection results for the setting in \eqref{eq:alter1}
are presented in Table~\ref{tab:alter1}. It can be seen that in all three
settings, the MAB, MSD and MMSE are all larger than when there is no spatially
varying covariate effects. There have been decrease in the MCR for parameters
corresponding to covariates that are in the true model. The variable selection
procedure, however, remains quite robust, and in all replicates of simulation
are the correct models selected. The average bandwidths selected for the three
settings are around 67. This indicates that in the presence of spatial
variation, for some replicates of simulation, the bandwidths tend to be large as
the gain in stabilizing the parameter estimates for each location dominates the
incurred bias. This has also been observed for the classical frequentist GWR, as
presented in Supplemental Table~2. The frequentist GWR
has an average effective number of parameters of value 19.32, 20.49 and 20.19
under the three settings, and the Bayesian GWR has 179.38, 179.58 and 179.32
instead.

\begin{table}[tbp]
\centering
\caption{Performance of parameter estimates, and variable selection results when
a simple linear variation pattern is present. ``True" denotes whether a
covariate is in the true model or not, and  ACC stands for variable selection
accuracy rate.}\label{tab:alter1}
    \begin{tabular}{cccccccccccc}
    \toprule
    & & MAB & MSD & MMSE & MCR & True & ACC(\%) & Model
ACC(\%) & $b$ \\
    \midrule 
Setting 1 & $\beta_1$  & 0.107 & 0.080 & 0.017 & 0.778 & 1 & 100 & 100 &
68.447\\
    & $\beta_2$  & 0.041 & 0.050 & 0.002 & 0.998 & 0 & 100 \\
    & $\beta_3$  &  0.040 & 0.051 & 0.003 & 0.980 & 0 & 100 \\
    & $\beta_4$  &  0.111 & 0.088 & 0.019 & 0.761 & 1 & 100 \\
    & $\beta_5$  & 0.111 & 0.087 & 0.019 & 0.761 & 1 & 100\\ [0.5ex]
    Setting 2 & $\beta_1$  & 0.107 & 0.080 & 0.017 & 0.786 & 1 & 100 & 100 &
    67.677\\
    & $\beta_2$  & 0.111 & 0.088 & 0.019 & 0.766 & 1 & 100 \\
    & $\beta_3$  &  0.040 & 0.050 & 0.003 & 0.980 & 0 & 100\\
    & $\beta_4$  & 0.111 & 0.089 & 0.019 & 0.766 & 1 & 100 \\
    & $\beta_5$  & 0.111 & 0.086 & 0.018 & 0.764 & 1 & 100 \\ [0.5ex] Setting
    3
    &
    $\beta_1$  &  0.107 & 0.081 & 0.018 & 0.786 & 1 & 100 & 100 & 66.920\\
    & $\beta_2$  & 0.112 & 0.089 & 0.019 & 0.772 & 1 & 100 \\
    & $\beta_3$  & 0.110 & 0.084 & 0.018 & 0.777 & 1 & 100 \\
    & $\beta_4$  &  0.112 & 0.090 & 0.019 & 0.766 & 1 & 100 \\
    & $\beta_5$  & 0.111 & 0.087 & 0.019 & 0.767 & 1 & 100 \\
    \bottomrule
    \end{tabular}
\end{table}

\begin{figure}[tbp]
\centering
    \includegraphics[width=\textwidth]{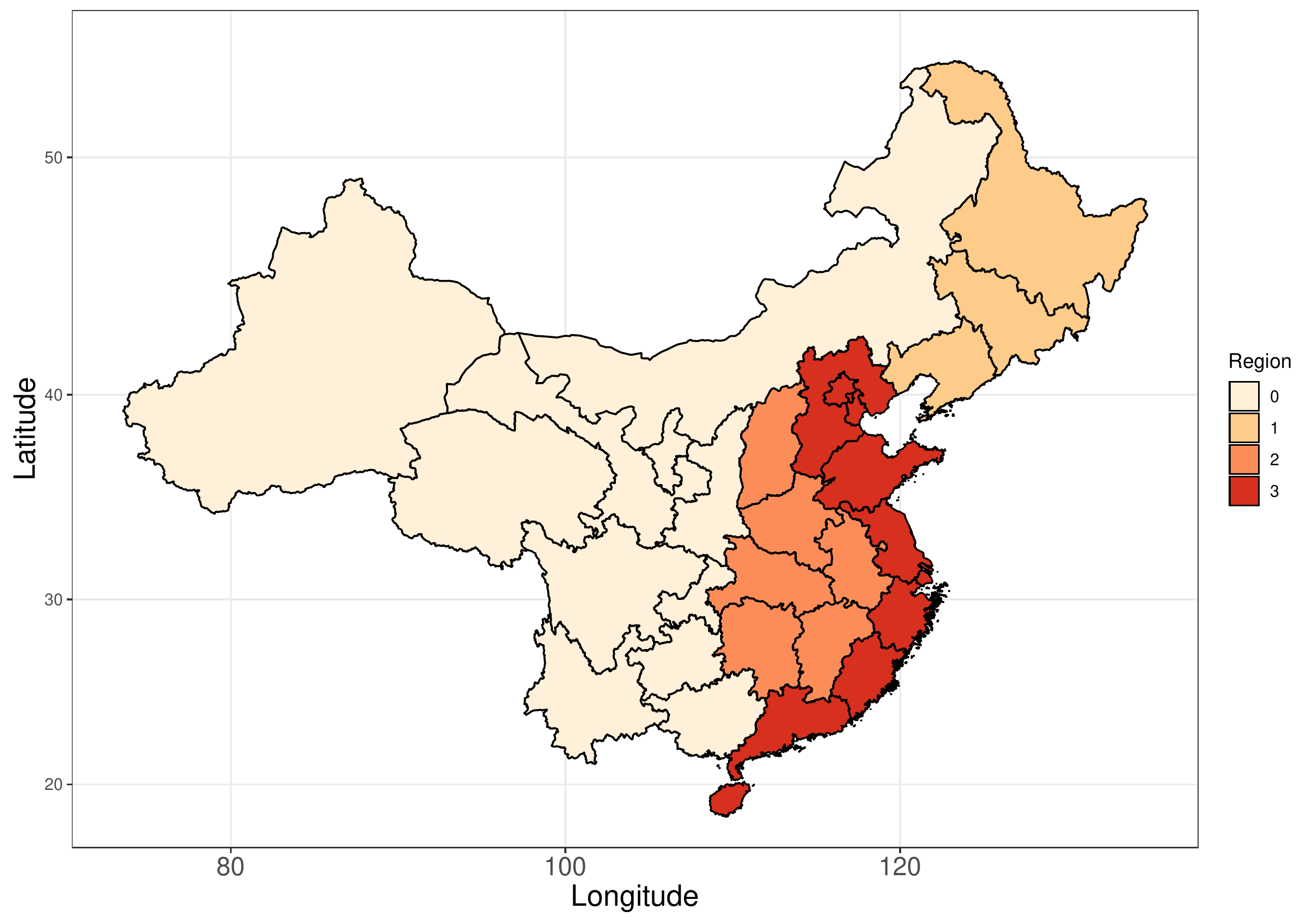}
    \caption{Map of selected provinces in China colored by their economic
regions proposed during the Eleventh Five-year Plan.}
    \label{fig:regions}
\end{figure}

To study the estimation and variable selection performance under a scenario
where regional variation exists, we choose to use the four major economic
regions of China proposed during the Eleventh Five-year plan: the west (0),
northeast (1), central (2), and east (3) regions. A visualization of these four
regions is given in Figure~\ref{fig:regions}. Provinces within each economic
region are assigned the same parameter value. The four $\beta^\top$'s under the
three simulation settings are given in Table~\ref{tab:truebetas}. The estimation
and variable selection results are presented in Table~\ref{tab:alterregion}.
Again, compared to results in Supplemental Table~3, both
approaches yield similar MAB, MSD, MMSE and MCR/MCP. Similar to previously
observed, the variable selection in the Bayesian approach effectively reduces
the MAB, MSD and MMSE parameter estimates for variables that are not in the true
underlying model. The bandwidths selected average to around 70. This could be
due to the fact that a province now has a few neighbors with exactly the same
true underlying coefficients, and therefore the weighting function tries to
weigh observations in the neighboring provinces equally as the local one. The
frequentist GWR performed similarly, and results are included Supplemental
Table~3. The average effective number of parameters under
the frequentist GWR are 17.27, 16.73 and 15.28 under the three settings, while
under the Bayesian GWR, the values are 178.35, 178.25 and 178.38, respectively.

\begin{table}[tbp]
    \centering 
    \caption{The true underlying parameters used for the four regions in each
setting.}\label{tab:truebetas}
    \begin{tabular}{lllll}
        \toprule
        & Region 0 & Region 1 & Region 2 & Region 3 \\ \midrule Setting 1 &
(1.8, 0, 0, 4.2, 7) & (1.5, 0, 0, 3.8, 9) & (2.2, 0, 0, 4, 8.5) & (2, 0,
        0, 4, 8)\\
        Setting 2 & (1.8, 1.8, 0, 4.2, 7) & (1.5, 1.5, 0, 3.8, 9) & (2.2, 2.2,
        0, 4, 8.5) & (2, 2, 0, 4, 8)\\
        Setting 3 & (1.8, 1.8, 2.9, 4.2, 7) & (1.5, 1.5, 3.4, 3.8, 9) & (2.2,
        2.2, 3.1, 4, 8.5) & (2, 2, 3, 4, 8)\\
        \bottomrule 
    \end{tabular}
\end{table}

\begin{table}[tbp]
\centering
\caption{Performance of parameter estimates and variable selection results when
regional variation is present. ``True'' denotes whether a covariate is in the
true model or not, and ACC stands for variable selection accuracy
rate.}\label{tab:alterregion}
	\begin{tabular}{cccccccccccc}
    \toprule
    & & MAB & MSD & MMSE & MCR & True & ACC(\%) & Model
    ACC(\%) & $b$ \\
    \midrule 
    Setting 1 & $\beta_1$ & 0.205 & 0.107 & 0.061 & 0.557 & 1 & 100 & 100 &
    69.469\\
    & $\beta_2$ & 0.041 & 0.049 & 0.002 & 1.000 & 0 & 100 \\
    & $\beta_3$ & 0.043 & 0.053 & 0.003 & 0.999 & 0 & 100 \\
    & $\beta_4$ & 0.136 & 0.111 & 0.028 & 0.782 & 1 & 100 \\
    & $\beta_5$ & 0.662 & 0.125 & 0.554 & 0.167 & 1 & 100 \\ [0.5ex] 
    Setting 2 &
    $\beta_1$ & 0.205 & 0.108 & 0.061 & 0.555 & 1 & 100 & 100 & 70.156 \\
    & $\beta_2$ & 0.204 & 0.105 & 0.061 & 0.578 & 1 & 100 \\
    & $\beta_3$ & 0.044 & 0.054 & 0.003 & 0.999 & 0 & 100 \\
    & $\beta_4$ & 0.136 & 0.114 & 0.029 & 0.791 & 1 & 100 \\
    & $\beta_5$ & 0.662 & 0.126 & 0.554 & 0.168 & 1 & 100 \\
    [0.5ex]
    Setting 3 & $\beta_1$ & 0.205 & 0.111 & 0.062 & 0.559 & 1 & 100 & 100 &
    70.491\\
    & $\beta_2$ & 0.205 & 0.106 & 0.061 & 0.589 & 1 & 100 \\
    & $\beta_3$ & 0.146 & 0.113 & 0.034 & 0.743 & 1 & 100 \\
    & $\beta_4$ & 0.138 & 0.116 & 0.029 & 0.790 & 1 & 100 \\
    & $\beta_5$ & 0.662 & 0.126 & 0.554 & 0.171 & 1 & 100 \\
    \bottomrule
    \end{tabular}
\end{table}

\section*{Real Data Analysis}\label{sec:realdata}

\begin{figure}[tbp]
    \centering
    \includegraphics[width=\textwidth]{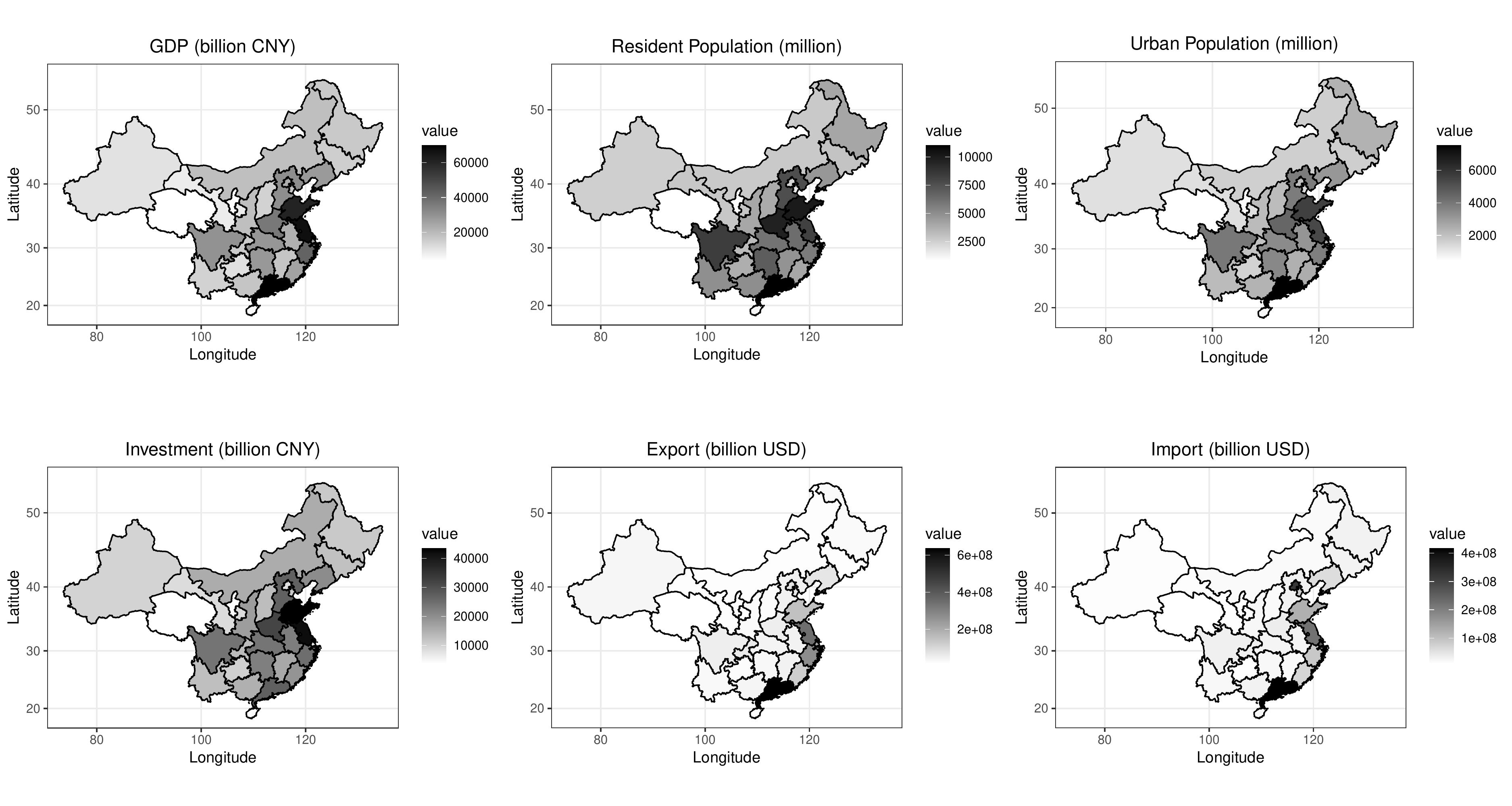}
    \caption{Five-year means for variables in selected provinces of China.}
    \label{fig:varplot}
\end{figure}

The proposed Bayesian GWR model is used to analyze province-level macroeconomic
data in 30 selected provinces of China from year 2012 to year 2016, i.e., we
have 150 observations in total. The Gross Domestic Product (GDP, in billions of
CNY) is used as the spatial response variable~($Y$). Five covariates, including
the resident population in millions~($X_1$), the urban population in
millions~($X_2$), the fixed asset investment in the whole society in billions of
CNY~($X_3$), total export value in billions of USD~($X_4$), and total import
value in billions of USD~($X_5$), are incorporated in the model. The 5-year
means of the variables for each province are shown in Figure~\ref{fig:varplot}.
Following the common practice in economics to account for long-tails \citep[see,
e.g.][]{wooldridge2015introductory}, we take the logarithm of GDP before model
fitting. All five covariates are continuous, and are therefore standardized
before model fitting.

The proposed Bayesian GWR model \eqref{eq:Ymodel} -- \eqref{eq:bmodel} is fitted
on this dataset. Priors $\sigma^2(s) \sim \text{IGamma}(1,1)$ and $\beta_0(s)
\sim N(0,1)$ are given, and we set $\tau_j^2 = 0.001$, $c_j^2=10000$ following
the common practice in spike-and-slab model selection, and $D = 100$ so that we
start from an approximately uniform weight over all provinces. The same graph
distance matrix as in Section~\ref{sec:simu} was used. The length of chains was
selected to be 5000, with the first 2000 as burn-in. The unity, exponential, and
Gaussian weighting schemes were considered, and the DIC and LPML were used to
select the best among the three for this particular dataset. The DIC and LPML
values as well as the effective number of parameters
\citep[$p_D$;][]{spiegelhalter2002bayesian}
for these three weighting schemes are shown in Table~\ref{tab:DICLPML}. It can
be seen that the Gaussian weighting scheme yields the smallest DIC value and the
largest LPML value, indicating that the model with a Gaussian weighting scheme
is selected as the best model among the candidate models.

\begin{table}[tbp]
	\centering
\caption{DIC and LPML values for different weighting
schemes}\label{tab:DICLPML}
		\begin{tabular}{lccc}
			\toprule
			& Unity Scheme & Exponential Scheme & Gaussian Scheme \\
			\midrule 
			DIC & 12584.45 & 12525.65  & 12511.21 \\
			LPML & -6875.99 & -6876.55 & -6871.79 \\
		    $p_D$ & 214.47 & 180.68 & 190.00 \\
			\bottomrule
		\end{tabular}
\end{table}

\begin{figure}[tbp]
    \centering
    \includegraphics[width=\textwidth]{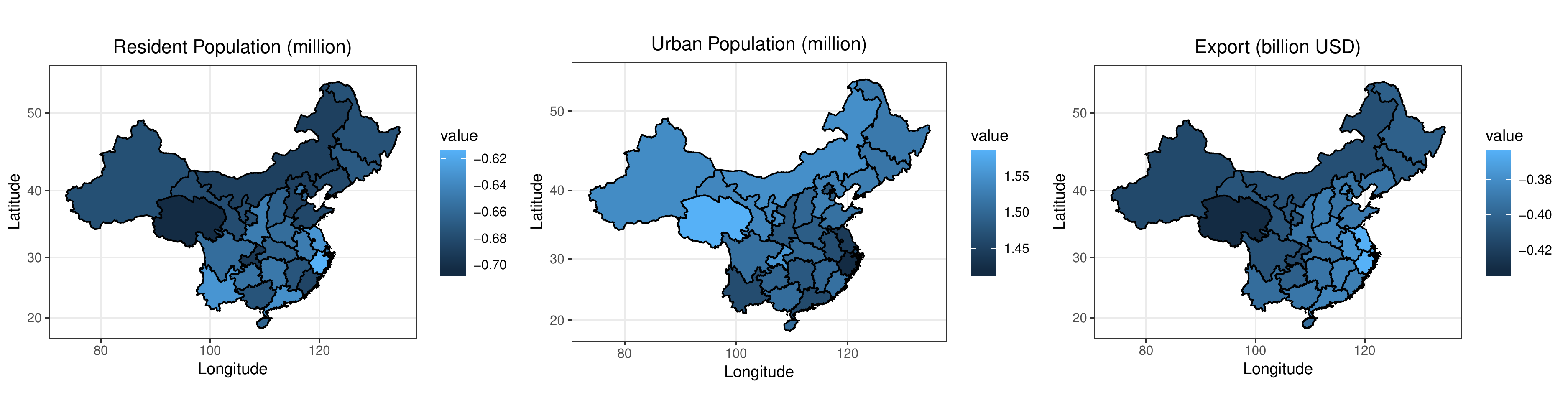}
    \caption{Plot of parameter estimates for the three covariates in the final
model in each selected province.}\label{fig:finalbetahat}
\end{figure}

Under the GWR model with a Gaussian weighting scheme, the posterior modes of the
indicators~$\gamma_j$ are $(1,1,0,1,0)$, respectively, which indicates that the
covariates $X_1$, $X_2$ and $X_4$ are selected, while covariates $X_3$ and $X_5$
can be excluded from the regression model. Specifically, in our model, the
number of resident population, the number of urban population, and total export
value can help explain the change of GDP in each province. The posterior
estimation results of the parameters under the Bayesian GWR model with a
Gaussian weighting scheme are presented in Figure~\ref{fig:finalbetahat}. The
geographical variation in the parameters is rather obvious. We can see that the
number of resident population and the total export value have significant
negative impact on the increase of GDP, while the number of urban population has
significant positive impact. For the Gaussian weight function, the posterior
estimate of the bandwidth $b$ is 9.40, which indicates that the most distant
provinces are assigned a relative weight of 0.665 in the estimation for one
particular province. The impact of resident population on GDP is larger in north
China than in southeast China. Comparing this to the population density map in
China made by the Center for Geographic Analysis at Harvard University
(\url{worldmap.havard.edu/maps/11756}), it can be seen that the influence is
bigger in less populous provinces, which is in accordance with our intuition.
The increasing effect of urban population on GDP is smaller in southeast China
than in northwest China. Considering the relatively higher urbanization in south
and east China \citep{Wang2012}, this can be explained by the ``decreasing
marginal effect" in economics. Export appears to be more important to provinces
in west China than in eastern areas, which can also be explained by the fact
that east China is relatively more developed, and have a more diversified source
of GDP.

For comparison, classical frequentist GWR in \eqref{eq:freqGWR} is also fitted
on the dataset. We report the parameter estimates, together with plots, in the
supplemental material. Particularly, the two covariates dropped by our Bayesian
variable selection approach have the smallest absolute parameter values among
all five covariates, which indicates that our proposed approach is indeed
capable of picking out the most influential factors. The parameter estimates are
also plotted on maps as in Figure~\ref{fig:finalbetahat}. There are slight
differences in the values of parameter estimates, which is partly due to the
fact that we only have 150 observations (30 provinces, 5 years). It is, however,
worth noticing that the trend of variation is consistent between the frequentist
and Bayesian approaches.

\section*{Discussion}\label{sec:discuss}

We developed a likelihood-based Bayesian approach to estimate regression
coefficients in conjunction with spike and slab variable selection for
geographically sparse data. The selection of bandwidth is discussed for a wide
choice of weighting schemes using popular Bayesian model selection criteria such
as the DIC and the LPML under the GWR context. The proposed methods are
implemented in \pkg{nimble}. In our simulation studies, when there is no
spatially varying covariate effect, the bandwidth is selected to give all
observations close to uniform weight in estimating the coefficients for each
individual location, whereas when there is indeed spatially varying covariate
effect, the bandwidth is selected to achieve a balance between introducing bias
for each location by taking into consideration nearby observations, and having
too unstable estimates by placing the majority of emphasis on local observations
and weighing down all others too heavily. The parameters estimated for each
location have decent coverage rate that are close to the nominal 95\% level.


\blue{Compared to the great circle distance, with a natural threshold of~1
to define ``close enough'', the graph distance yields weighting schems that
produce models with robust parameter estimation and variable selection
performance.}
Based on comparisons with classical frequentist GWR from the simulation studies,
it is interesting for us to notice that the bandwidth selection results of the
Bayesian approach are different from that of the frequentist approach. A partial
reason for this pattern is that the bandwidth selection of frequentist approach
is based on minimizing the summation of the sum of square errors (SSE) for each
locations. Our Bayesian approach, however, tries to maximize the whole data
likelihood. Therefore, the frequentist approach tends to select smaller
bandwidth than Bayesian approach.

A few issues beyond the scope of this paper are worth further investigation. In
this work, we are only concerned with estimation of parameter for linear
regression. Extension of similar ideas to generalized linear models, and
semi-parametric models such as the Cox model, are worth developing. In the
second alternative simulation scenario with regional variation patterns, both
the frequentist and Bayesian GWR try to weigh neighbors as high as possible,
leading to large bandwidths. Under the frequentist framework, clustering of
covariate effects have been done using hierarchical clustering on the parameter
estimation, which is ad hoc. Another approach is the penalized methods in
\cite{li2019spatial}. In the Bayesian paradigm, however, hierarchical modeling
provides an integrated framework that incorporates the latent cluster
configuration layer. Development of such a framework is worth investigating.
Also, we are assuming that a covariate is either in the true model for all
locations, or not in the true model for all locations. There are cases where a
covariate is important for some locations, but is minimally impacting for other
locations. Identifying such locations is devoted to future research. Detecting a
relationship between two areas that do not share a boundary
\citep{gao2019bayesian} other than using graph distance is also an interesting
future work.

\end{document}